\documentclass[prl,twocolumn,showpacs,superscriptaddress,floatfix,
preprintnumbers,nofootinbib]{revtex4}

\usepackage{epsfig}
\usepackage{amsmath,bm}
\usepackage{hyperref}
\usepackage{color}

\begin{document}

\title{Neutrino signature of supernova hydrodynamical instabilities in three dimensions}

\author{Irene Tamborra}
\affiliation{Max-Planck-Institut f\"ur Physik (Werner-Heisenberg-Institut),
 F\"ohringer Ring 6, 80805 M\"unchen, Germany}

\author{Florian Hanke}
\affiliation{Max-Planck-Institut f\"ur Astrophysik,
 Karl-Schwarzschild-Str.~1, 85748 Garching, Germany}

\author{Bernhard~M\"uller}
\affiliation{Max-Planck-Institut f\"ur Astrophysik,
 Karl-Schwarzschild-Str.~1, 85748 Garching, Germany}

\author{Hans-Thomas~Janka}
\affiliation{Max-Planck-Institut f\"ur Astrophysik,
 Karl-Schwarzschild-Str.~1, 85748 Garching, Germany}

\author{Georg Raffelt}
\affiliation{Max-Planck-Institut f\"ur Physik (Werner-Heisenberg-Institut),
 F\"ohringer Ring 6, 80805 M\"unchen, Germany}

\date{\today}

%%%%%%%%%%%%%%%%%%%%%%%%%%%%%%%%%%%%%%%%%%%%%%%%%%%%%%%%%%%%%%%%%%%

\begin{abstract}
The first full-scale three-dimensional (3D) core-collapse supernova
(SN) simulations with sophisticated neutrino transport show
pronounced effects of the standing accretion shock instability
(SASI) for two high-mass progenitors (20 and $27\,M_\odot$). In a
low-mass progenitor ($11.2\,M_\odot$), large-scale convection is the
dominant nonradial hydrodynamic instability in the postshock
accretion layer. The SASI-associated modulation of the neutrino
signal (80\,Hz in our two examples) will be clearly detectable in
IceCube or the future Hyper-Kamiokande detector, depending on
progenitor properties, distance, and observer location relative to
the main SASI sloshing direction. The neutrino signal from the next
galactic SN can, therefore, diagnose the nature of the hydrodynamic
instability.
\end{abstract}

\preprint{MPP-2013-191}


\maketitle

%%%%%%%%%%%%%%%%%%%%%%%%%%%%%%%%%%%%%%%%%%%%%%%%%%%%%%%%%%%%%%%%%%%%%%
% Introduction
%%%%%%%%%%%%%%%%%%%%%%%%%%%%%%%%%%%%%%%%%%%%%%%%%%%%%%%%%%%%%%%%%%%%%%

{\em Introduction}.---Bethe and Wilson's delayed neutrino-driven
explosion mechanism~\cite{Bethe:1985} remains the standard
core-collapse SN paradigm~\cite{Janka:2012wk}. At core bounce a shock
wave forms, stalls after reaching 100--200\,km, and is revived by
neutrino heating after tens to hundreds of ms, depending on
progenitor properties and accretion rate of stellar matter that
continues to collapse. One modern key ingredient to this scenario is
its inherent multi-dimensional nature inferred from observed SN
asymmetries~\cite{Arnett:1989} and from parametric and
self-consistent 2D and 3D hydrodynamical
simulations~\cite{Herant:1994, Burrows:1995, Janka:1996, Fryer:2002}.
During the accretion phase, large-scale convective overturn develops
in the neutrino-heated postshock layer~\cite{Bethe:1990mw} and the
standing accretion shock instability (SASI) can arise, involving
global dipolar and quadrupolar deformation and sloshing motions of
the shock front~\cite{Blondin:2002sm, Scheck:2007gw} as well as
spiral modes~\cite{Blondin:2007, Iwakami:2009, Fernandez:2010,
Foglizzo:2011aa, Hanke:2013ena}. The next galactic SN may reveal
these effects in gravitational waves~\cite{Marek:2008qi, Kotake:2009,
Murphy:2009, Mueller:2013} and in neutrino flux
variations~\cite{Lund:2010kh, Lund:2012vm}.

Most SN investigations of convection and SASI have relied on
axisymmetric simulations where sloshing  motions are constrained to
the symmetry axis \cite{Marek:2008qi, Brandt:2010xa, Marek:2007gr,
Mueller:2012is, Mueller:2012ak, Murphy:2008dw, Nordhaus:2010uk}.
Several recent 3D models have treated neutrino heating and cooling in
the SN core in various approximations \cite{Iwakami:2007ie,
Wongwathanarat:2010ip, Muller:2011yi, Hanke:2011jf, Takiwaki:2011db,
Burrows:2012yk, Ott:2013}. They found SASI sloshing motions with
considerably reduced amplitudes and stochastically changing direction
or no clear SASI signature at all. Buoyancy-driven convection was
concluded to dominate post-shock turbulence and SASI to be a minor
feature of SN dynamics at
best~\cite{Burrows:2012yk,Murphy:2012id,Dolence:2012kh}. However,
self-consistent, 2D, general relativistic simulations with
sophisticated neutrino transport suggest that a genuine SASI remains
possible if the shock stagnation radius is sufficiently
small~\cite{Mueller:2012ak}. SASI development may depend on both,
progenitor properties and the exact behavior of the stalled shock,
which requires reliable neutrino transport. So the importance of the
SASI relative to neutrino-driven convection remains controversial. Therefore
 it is remarkable that the first 3D simulation with detailed
neutrino transport (a 27\,$M_\odot$ model) shows violent SASI
activity~\cite{Hanke:2013ena}.

SASI activity strongly modulates the accretion flow to the neutron
star and the associated neutrino emission \cite{Marek:2008qi,
Brandt:2010xa}. The detection of such fast time variations of the
neutrino signal will offer a unique chance to probe stellar core
collapse and its detailed astrophysics~\cite{Lund:2010kh,
Lund:2012vm}. A significant signal must stick above the shot noise
caused by the fluctuating event rate. IceCube~\cite{Abbasi:2011ss,
Demiroers:2011am} is among the most promising facilities for this
task, detecting a large number of Cherenkov photons triggered by
neutrinos. Moreover, Super-Kamiokande (Super-K) \cite{Abe:2010hy}, or
the next-generation Hyper-Kamiokande (Hyper-K) \cite{Abe:2011ts},
although with smaller rate than IceCube, will monitor the neutrino
signal without background and will provide event-by-event energy
information. The once-in-a-lifetime opportunity to
observe a high-statistics SN neutrino signal provides one of several
physics motivations to build, maintain, and constantly improve such
large neutrino observatories.

We here study the detection opportunities for a SASI-modulated SN
neutrino signal based on the world-wide first 3D simulations with
detailed neutrino transport of three progenitors with
$27\,M_\odot$~\cite{Hanke:2013ena}, $20\,M_\odot$, and
$11.2\,M_\odot$. Whenever vigorous SASI motions grow despite
neutrino-driven convection, the neutrino signal modulations will be
clearly detectable for a galactic SN, but the exact signal features
depend on progenitor properties, SN distance, and location of the
observer relative to the main sloshing directions.

%%%%%%%%%%%%%%%%%%%%%%%%%%%%%%%%%%%%%%%%%%%%%%%%%%%%%%%%%%%%%%%%%%%%%%%
%SN models
%%%%%%%%%%%%%%%%%%%%%%%%%%%%%%%%%%%%%%%%%%%%%%%%%%%%%%%%%%%%%%%%%%%%%%%

{\em Numerical supernova models}.---We use solar metallicity
progenitors for which the evolution until the onset of iron-core
collapse has been reported in~\cite{Woosley:2002} for the 11.2 and
27\,$M_\odot$ stars and in~\cite{Woosley:2007} for the 20\,$M_\odot$
star. They were previously employed for 2D
simulations~\cite{Marek:2007gr, Mueller:2012is, Mueller:2012ak,
Bruenn:2013}. Our 3D modeling uses the {\textsc{Prometheus-Vertex}}
hydrodynamics code. It includes a ``ray-by-ray-plus'' (RbR+), fully
velocity and energy-dependent neutrino transport module based on a
variable Eddington-factor technique that solves iteratively the
neutrino energy, momentum, and Boltzmann equations \cite{Rampp:2002,
Buras:2006}. We employ state-of-the-art neutrino interaction
rates~\cite{Buras:2006, Mueller:2012is} and relativistic gravity and
redshift corrections~\cite{Rampp:2002, Marek:2006}.

The RbR+ description assumes the neutrino momentum distribution to be
axisymmetric around the radial direction everywhere, implying that
the neutrino fluxes are radial. The detectable energy-dependent
neutrino emission from the hemisphere facing an observer is
determined with a post-processing procedure that includes projection
and limb-darkening effects~\cite{Muller:2011yi}. We will use the
$27\,M_\odot$ model as our benchmark case because its properties have
been published~\cite{Hanke:2013ena}. Details of the other two
simulations will be provided elsewhere~\cite{Tamborra:2013prep}.
 All simulations used artificial random density
perturbations of 0.1\% amplitude on the whole numerical grid to seed
the growth of hydrodynamic instabilities. None of the models had
exploded at the end of the computation runs.

%%%%%%%%%%%%%%%%%%%%%%%%%%%%%%%%%%%%%%%%%%%%%%%%%%%%%%%%%%%%%%%%%%%%%%%
%IceCube
%%%%%%%%%%%%%%%%%%%%%%%%%%%%%%%%%%%%%%%%%%%%%%%%%%%%%%%%%%%%%%%%%%%%%%%

{\em Detector signal}.---In the largest operating detectors, IceCube
and Super-K, neutrinos are primarily detected by inverse beta decay,
$\bar\nu_e+p\to n+e^+$, through Cherenkov radiation of the positron.
We represent the neutrino emission spectra in the form of Gamma
distributions \cite{Keil:2002in, Tamborra:2012ac}. We estimate the
neutrino signal following the IceCube Collaboration
\cite{Abbasi:2011ss}, accounting for a $\sim$13\% dead-time effect
for background reduction. We use a cross section that includes
recoil effects and other corrections~\cite{Strumia:2003zx}, overall
reducing the detection rate by  30\% relative to earlier
studies~\cite{Dighe:2003jg, Lund:2010kh, Lund:2012vm}. On the other
hand, we increase the rate by 6\% to account for detection channels
other than inverse beta decay \cite{Abbasi:2011ss}.

We assume an average background of $0.286~{\rm ms}^{-1}$ for each of
the 5160 optical modules, i.e., an overall background rate of
$R_{\mathrm{bkgd}}=1.48\times10^3~\mathrm{ms}^{-1}$, comparable to
the signal rate for a SN at 10~kpc. The IceCube data acquisition
system has been upgraded since the publication of
Ref.~\cite{Abbasi:2011ss} so that the full neutrino time sequence
will be available instead of time bins.

IceCube will register in total around $10^6$ events above background
for a SN at 10~kpc, to be compared with around $10^4$ events for
Super-K (fiducial mass 32~kton), i.e., IceCube has superior
statistics. On the other hand, the future Hyper-K will have a
fiducial mass  of 740~kton, providing a background-free signal of
roughly 1/3 the IceCube rate. Therefore, Hyper-K can have superior
signal statistics, depending on SN distance. In addition, it has
event-by-event energy information which we do not use for our simple
comparison.

\begin{figure}
\includegraphics[width=0.9\columnwidth]{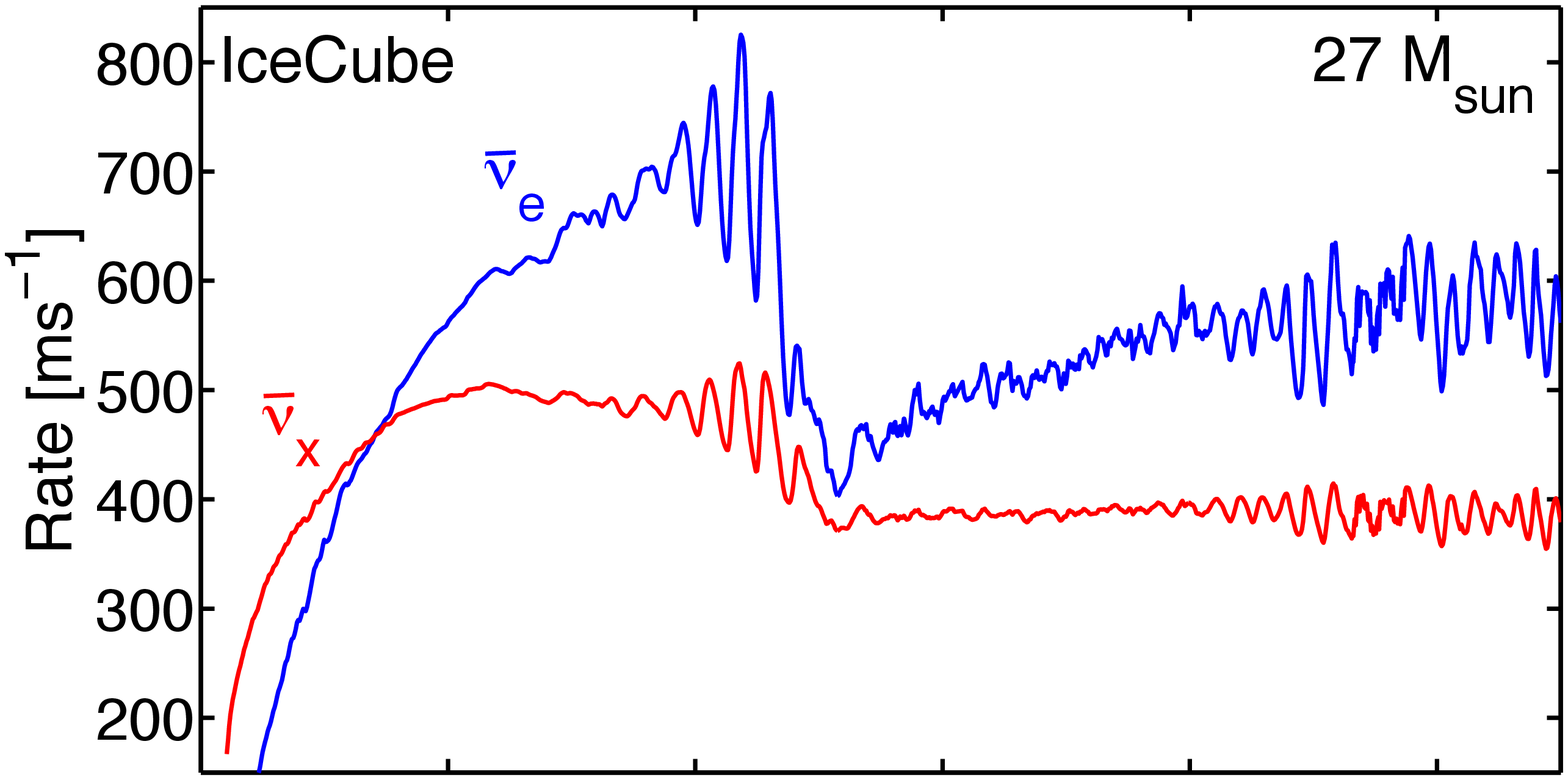}\\
\includegraphics[width=0.9\columnwidth]{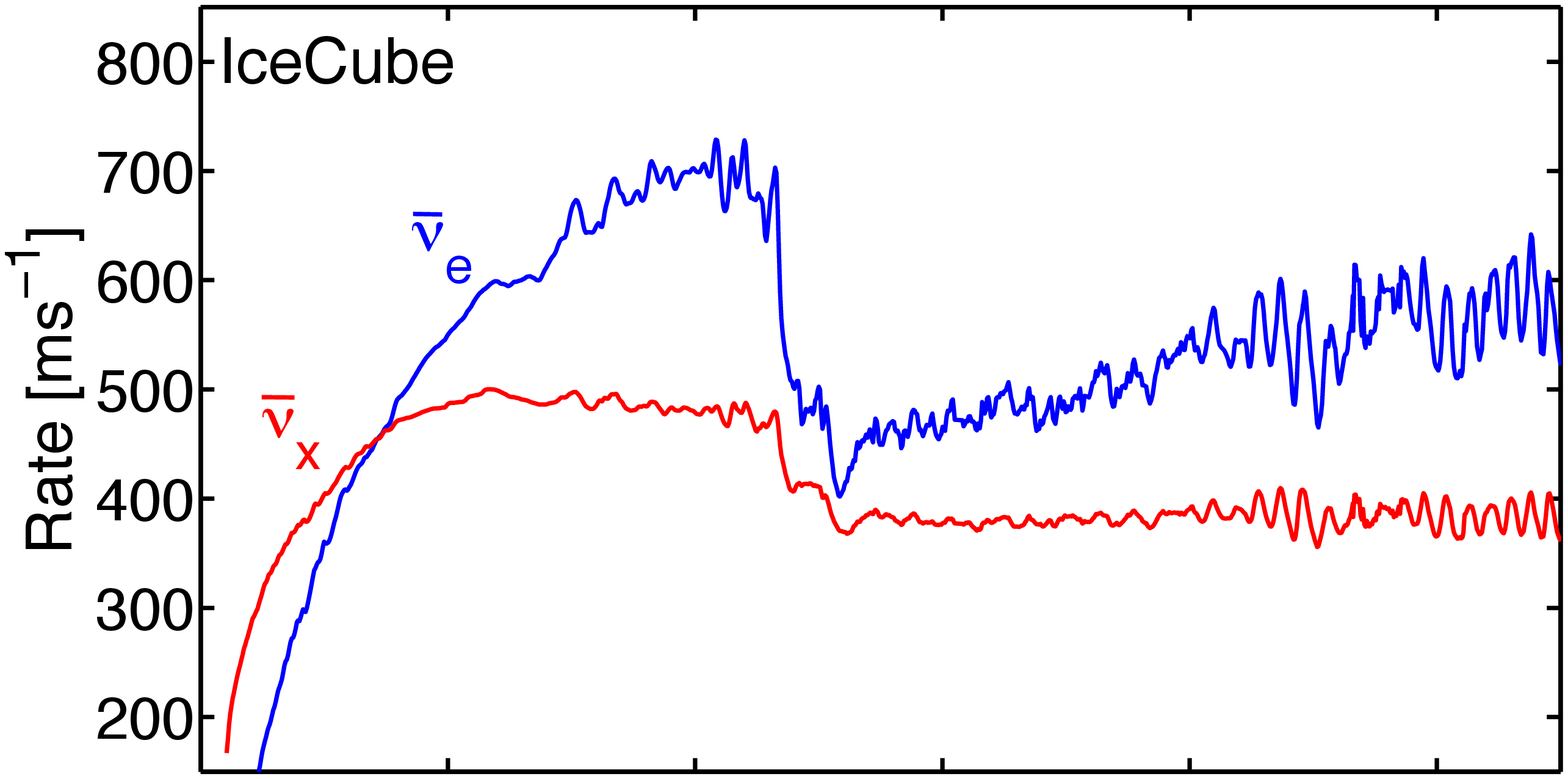}\\
\hspace{-1.5mm} \includegraphics[width=0.905\columnwidth]{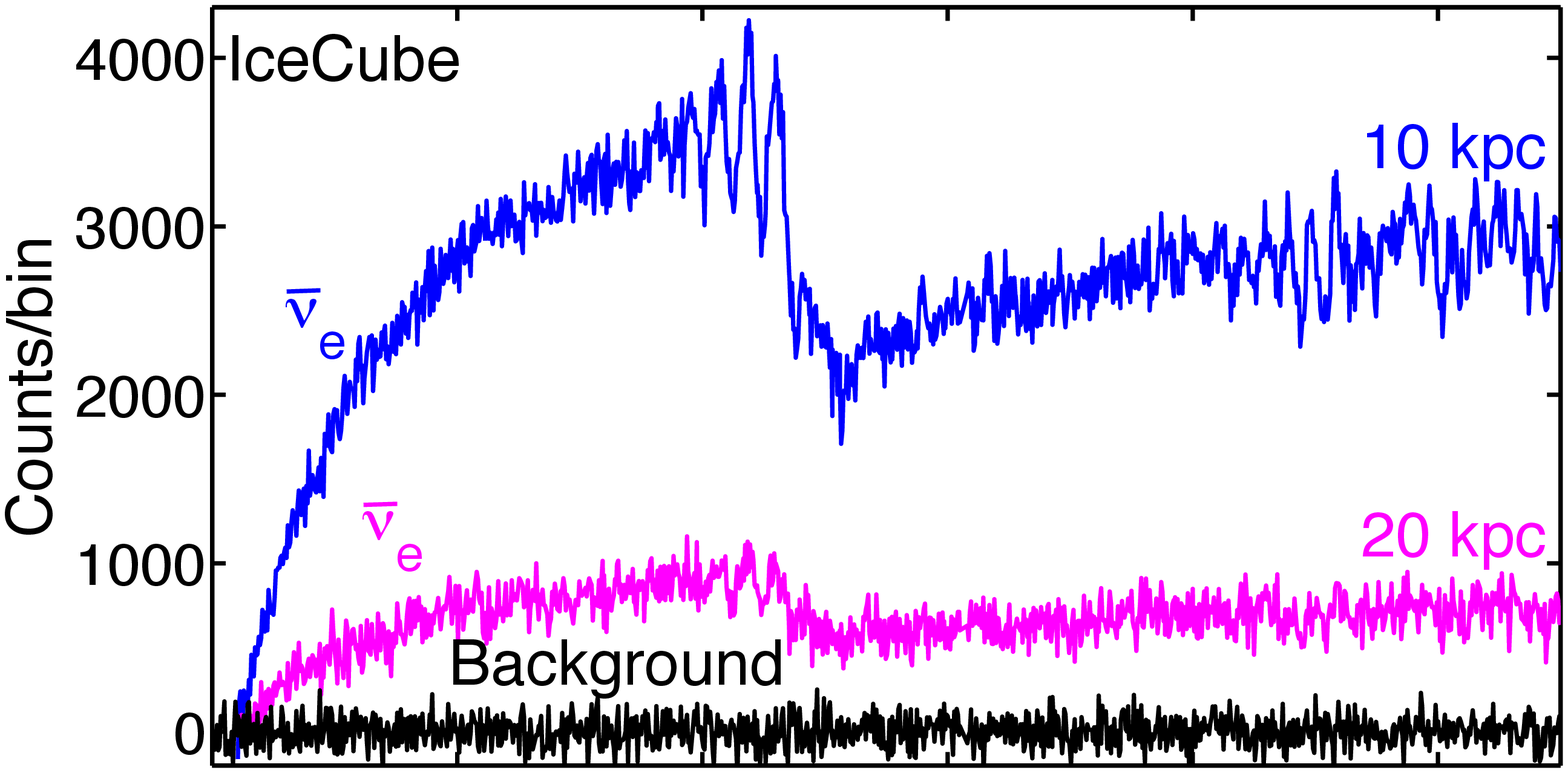}\\
\includegraphics[width=0.91\columnwidth]{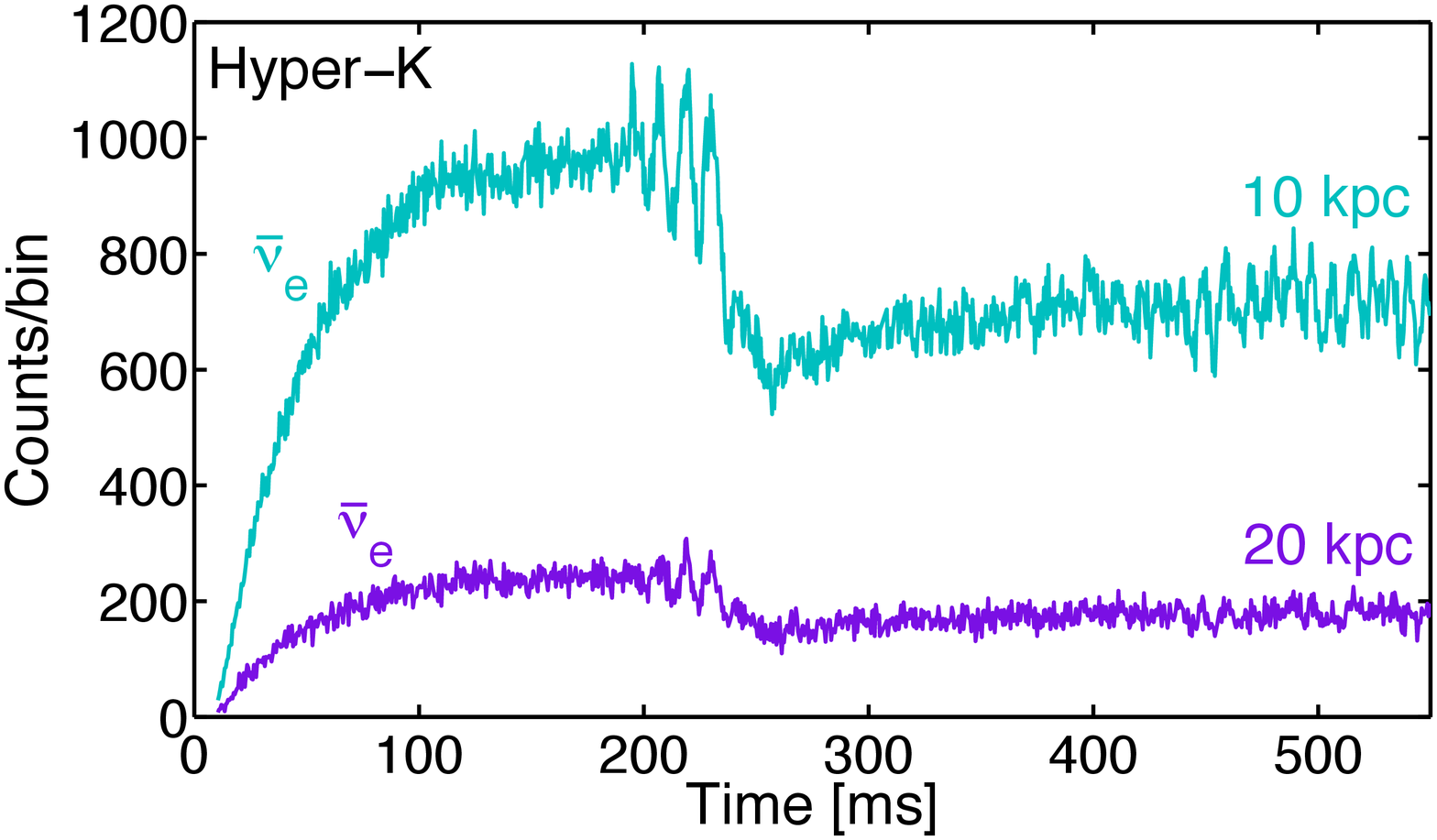}
\caption{Detection rate for our $27\ M_\odot$ SN progenitor,
upper panels for IceCube, bottom one for Hyper-K.
The observer direction is chosen for strong signal modulation,
except for the second panel (minimal modulation).
Upper two panels: IceCube rate at 10\,kpc for
$\bar\nu_e$ (no flavor conversion) and for $\bar\nu_x$
(complete flavor conversion).
The lower two panels include a random shot-noise realization,
5\,ms bins, for
the indicated SN distances. For IceCube also the background
fluctuations without a SN signal are shown.\label{fig:27Msun}}
\end{figure}

{\em Signal modulation in the $27\,M_\odot$ model}.---To get a first
impression of the  neutrino signal modulation we consider
our published $27\,M_\odot$ model~\cite{Hanke:2013ena}, meanwhile
simulated until $\sim$550~ms. This model shows clear SASI activity
at 120--260\,ms. At $\sim$220\,ms a SASI spiral mode
sets in and remains largely confined to an almost stable plane,
which is not aligned with the polar grid of the simulation. We
select an observer in this plane in a favorable direction
and show the expected IceCube signal in the top panel of
Fig.~\ref{fig:27Msun}. One case assumes the signal to be caused
by anti-neutrinos emitted as $\bar\nu_e$ at the source, i.e., we
ignore flavor conversions. The other case takes into account complete flavor
conversion so that the signal is caused by $\bar\nu_x$, i.e., a
combination of $\bar\nu_\mu$ and $\bar\nu_\tau$. Both cases reveal
large signal modulations with a clear periodicity.

The first SASI episode ends abruptly with the accretion of the
Si/SiO interface, followed by large-scale convection with much
smaller and less periodic signal modulations (see also Figs.~1, 2,
and 6 of Ref.~\cite{Hanke:2013ena}). After about 410~ms, SASI
activity begins again until the end of our simulation. The signal
modulation is now weaker, partly owing to a lower SASI amplitude and
partly to the chosen observer direction being no longer optimal.

The second panel of Fig.~\ref{fig:27Msun} is for a direction
orthogonal to the plane of the first SASI episode, i.e., the signal
modulation is particularly small. The second SASI episode now shows
a stronger signal than the first because the observer is no longer
in the worst direction.

\begin{figure}
\includegraphics[width=0.95\columnwidth]{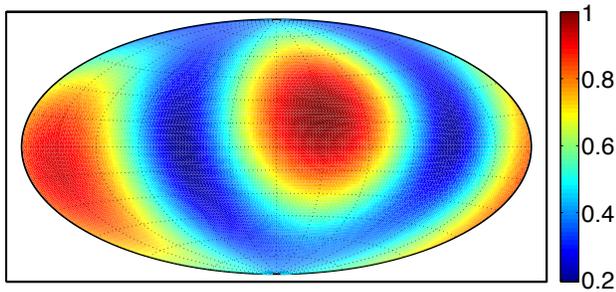}
\caption{Relative amplitude of the $\bar\nu_e$ rate modulation
(see Eq.~\ref{eq:sigma}) on a sky-plot of observer directions
during the first SASI episode (120--250\,ms) of the $27\,M_\odot$ model.
\label{fig:map_27}}
\end{figure}

The SASI sloshing and spiral motions imply that observers in
opposite directions obtain almost the same signal modulations with
opposite phase. To illustrate the dependence on the observer
direction we provide a supplementary  
\href{http://www.mpa-garching.mpg.de/ccsnarchive/data/Hanke2013_movie/index.html}{movie} on the time-evolution of
the IceCube rate~\footnote{$\mathtt{http}$://$\mathtt{www.mpa}$-$\mathtt{garching.mpg.de/ccsnarchive/data/}\newline\mathtt{Hanke2013\underline{\hspace{1.5mm}}movie/index.html}.$}. As
a static visualization we show in
Fig.~\ref{fig:map_27} the relative amplitude of the IceCube
detection rate during the first SASI episode. To define this
amplitude we first note that the signal rate, averaged over all
directions, hardly shows any modulation at all. In a given direction
we define the relative time-dependent rate and consider its root
mean square deviation for the first SASI episode
($[t_1,t_2]=[120,250]$\,ms),
\begin{equation}
\sigma \equiv \left ( \int_{t_1}^{t_2} dt\,
\left [
\frac{R - \left\langle R\right\rangle}{\left\langle R\right\rangle}
\right ]^{2} \right )^{\! 1/2} \,.
\label{eq:sigma}
\end{equation}
Despite the spiral mass motions during
this SASI episode and the corresponding, considerable time
variability of the emission asymmetry, the time integrated analysis still
reveals a dominant sloshing direction, which
produces two signal ``hot spots'' in two opposite directions,
surrounded by directions with much smaller modulations.

\begin{figure}
\includegraphics[width=0.9\columnwidth]{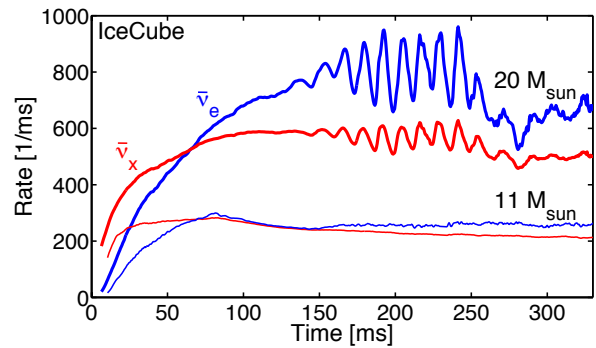}
\caption{IceCube rate for optimal observing directions for the
11.2 and $20\,M_\odot$ models at 10\,kpc, as in the  top
panel of Fig.~\ref{fig:27Msun}.\label{fig:11_20Msun}}
\end{figure}

{\em Other progenitors}.---Figure~\ref{fig:11_20Msun} shows the
IceCube rate for the other progenitors ($11.2$ and $20\,M_\odot$) in
optimal observing directions. For the heavier case, a strong SASI
develops after 140~ms. Again a global SASI spiral mode largely
confined to a plane appears, lasting until $\sim$300\,ms close to
the end of our simulation. The signal modulations are even more
pronounced than for the $27\,M_\odot$ progenitor and the SASI phase
lasts slightly longer. In contrast, the $11.2\,M_\odot$ model
exhibits dominant activity by neutrino-driven convective overturn in
the postshock layer (manifesting itself in a highly time-variable
pattern of rising high-entropy bubbles and cooler downflows) without
any clear signs of large-amplitude coherent SASI motions. In this
case only very small, short-time signal fluctuations are visible for
a chosen observer direction as a consequence of non-stationary,
chaotically changing accretion anisotropies (similar to the cases
analyzed in Refs.~\cite{Muller:2011yi, Lund:2012vm}), although
significant directional differences of the $\bar\nu_e$ signal can
exist~\cite{Tamborra:2013prep}. The detection rate is also much
smaller because of a lower luminosity.

{\em Shot noise}.---The main limitation to observing signal
modulations are random fluctuations in the detected neutrino time
sequence. In the third panel of Fig.~\ref{fig:27Msun} we show the
IceCube $\bar\nu_e$ signal in 5\,ms bins, including a random shot
noise realization. The signal is roughly 700\,ms$^{-1}$ near maximum,
plus $1.48\times10^3\,\mathrm{ms}^{-1}$ background, i.e., roughly
$1.1\times10^4$ events per bin, causing a $\sim$3\% random
fluctuation of the signal itself where the average background
is subtracted. We also show the IceCube signal in the absence of a
SN, i.e., the background fluctuations alone. For a SN at 20\,kpc,
roughly the edge of the expected galactic SN distance distribution
\cite{Mirizzi:2006xx, Adams:2013ana}, the signal is still visible to
the naked eye, although the bin-to-bin fluctuation is now roughly
10\%.

In the bottom panel of Fig.~\ref{fig:27Msun}, we show the analogous
signal for Hyper-K, which has no background and thus yields roughly
900 events/bin. Its 3\% bin-to-bin random fluctuation is almost
identical to IceCube. Doubling the distance reduces the signal by
four, but as there is no dark current, the fluctuations grow to about
7\%, i.e., at this distance Hyper-K is superior. We
conclude that if the observer is located in an optimal direction,
SASI can be detected throughout the galaxy.

A serious strategy to filter such signal modulations from the noise
in less obvious cases is beyond the scope of our work. However, we
also illustrate the signal in terms of its Fourier power spectrum,
following Ref.~\cite{Lund:2010kh}. We select the time interval of
100--300\,ms, where SASI develops for our progenitors. With the
adopted signal duration of $\tau=200$\,ms, the spacing of the
discrete Fourier frequencies is $\delta f=1/\tau=5~{\rm Hz}$. We use
a Hann window function on our interval to reduce edge effects in the
Fourier transform. The minimum requirement for signal detection is
that the Fourier spectrum sticks above background. The average power
spectrum of a random signal sequence does not depend on frequency.
Therefore, the IceCube dark current is a natural baseline and we use
its power to normalize the signal power spectrum.

\begin{figure}
\includegraphics[width=0.95\columnwidth]{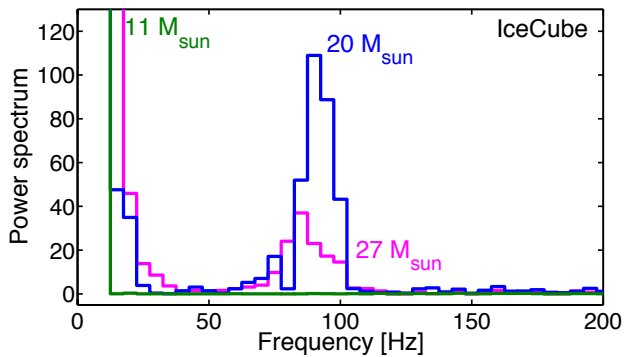}
\caption{Power spectrum of the IceCube event rate on the interval
100--300\,ms for our three progenitors, assuming the $\bar\nu_e$ signal
from a distance of 10\,kpc. Normalization is to the
frequency-independent power of shot noise caused by the IceCube
background of $1.48\times10^3\,\mathrm{ms}^{-1}$.\label{fig:ps_27}}
\end{figure}

Figure~\ref{fig:ps_27} shows the power spectrum of the IceCube event
rate for our three SN models thus normalized. A clear peak exists
at $\sim$80\,Hz for the two heavier progenitors where strong SASI
appears. The modulation frequency is determined by the variations of
the accretion flow that occur with the oscillation period of the
SASI mode. The corresponding SASI (fundamental) frequency $f_\mathrm{SASI}$
depends roughly
on the neutron star radius $R_\mathrm{NS}$ and
shock radius $R_\mathrm{S}$~\cite{Scheck:2007gw},
\begin{equation}
f_\mathrm{SASI}^{-1}\sim \int_{R_\mathrm{NS}}^{R_\mathrm{S}}
\frac{dr}{|v|} + \int_{R_\mathrm{NS}}^{R_\mathrm{S}}
\frac{dr}{c_\mathrm{s}-|v|} \, ,
\label{eq:fsasi}
\end{equation}
where $c_\mathrm{s}$ and $v$ are the radius-dependent sound
speed and accretion velocity, respectively, in the postshock layer.
For both the 27 and 20\,$M_\odot$ progenitors the SASI frequencies
are similar because of similar neutron star radii (the same equation
of state is used) and mean shock radii in the first 250\,ms after
bounce.

Note that the amplitude of the signal power spectrum varies with the
fourth power of distance. Therefore, doubling the distance to 20~kpc
reduces the power spectrum by a factor of 16 so that the main peak
of the $27\,M_\odot$ case reduces to about 3 times the shot-noise
level.

{\em Flavor oscillations}.---Neutrinos change their flavor as they
propagate from the emission region near the collapsed SN core to the
detector. None of the mixing angles is especially small so that the
propagation through the density gradient of the SN mantle and
envelope leads to adiabatic Mikheev-Smirnov-Wolfenstein (MSW)
conversions \cite{Mikheev:1986gs, Dighe:1999bi}. In particular, this
scenario predicts an approximately 70\% $\bar\nu_e$ survival
probability in the normal hierarchy of neutrino masses, whereas in
the inverted hierarchy, a complete swap
$\bar\nu_e\leftrightarrow\bar\nu_x$ is expected.

This traditional picture may be strongly modified by collective
flavor conversions between emission and the MSW region caused by
neutrino-neutrino refraction. Our theoretical understanding of this
effect is still developing. In the earlier literature one would have
predicted an almost complete flavor swap in the anti-neutrino
sector~\cite{Duan:2010bg}, reversing the hierarchy-dependence of the
above MSW predictions. Then it was recognized that the ordinary
matter effect can suppress collective flavor oscillations, especially
during the accretion phase of high-mass progenitors where the matter
density is large~\cite{EstebanPretel:2008ni, Chakraborty:2011nf,
Saviano:2012yh, Sarikas:2011am}. However, in those ``high-density''
cases the residual scattering of neutrinos beyond the emission region
is strong enough to provide a ``halo flux'' with backward angular
distribution, responsible for a significant modification of
neutrino-neutrino refraction~\cite{Cherry:2012zw}. This effect most
probably would not trigger collective flavor conversions in those
cases where the matter effect is important~\cite{Sarikas:2012vb}. The
latest development is the presence of yet another instability due to
the previously neglected azimuth variable of neutrino propagation,
whose matter suppression would require larger densities than
previously thought~\cite{Raffelt:2013rqa, RaffeltSeixas:2013,
Mirizzi:2013rla}.

In view of these unresolved complications, it is not clear which
exact flavor conversion scenario to expect as a function of the
neutrino mass hierarchy and depending on the progenitor properties.
For the purpose of detecting SASI-implied signal modulations, a
complete flavor swap where we observe $\bar\nu_e$ that were born as
$\bar\nu_x$ should be the worst case because of the smaller
modulation amplitude, although still detectable.

{\em Conclusions}.---The first sophisticated 3D SN simulations show
pronounced spiral SASI activity for the more massive of three
different progenitors. There are SASI phases interspersed with
episodes of dominant, large-scale convective overturn activity.
During the SASI periods, the neutrino signal modulations are even
larger than those seen in previous 2D simulations, whereas the
convective episodes are comparable to earlier 3D parametric
cases~\cite{Muller:2011yi,Lund:2012vm}. We have also shown that for
SN distances beyond some 10\,kpc, the future Hyper-K detector would
be superior to IceCube. In spite of its smaller signal rate (about
1/3 of IceCube), its lack of background implies a better
signal-to-noise ratio because of reduced shot noise for those
distances where IceCube is dominated by background fluctuations. To
exploit the full Hyper-K potential,  its event-by-event energy
determination should be used as well.

The neutrino signal of the next galactic SN, if captured by IceCube
and the future Hyper-K, offers a unique opportunity to diagnose
different types of hydrodynamical instabilities. Such detectable
instabilities appear in the first detailed 3D core-collapse SN
simulations and depend on the progenitor properties.

%%%%%%%%%%%%%%%%%%%%%%%%%%%%%%%%%%%%%%%%%%%%%%%%%%%%%%%%%%%%%%%%%%%%%%
%Acknowledgements %%%%%%%%%%%%%%%%%%%%%%%%%%%%%%%%%%%%%%%%%%%%%%%%%%%%
%%%%%%%%%%%%%%%%%%%%%%%%%%%%%%%%%%%%%%%%%%%%%%%%%%%%%%%%%%%%%%%%%%%%%%

{\em Acknowledgments.}---We thank Ewald M\"uller for discussions and
Nicole Schwarz for the animated visualization. This research was
partly supported by DFG through grants SFB/TR~7 and EXC~153 and by
the EU under grant PITN-GA-2011-289442 (FP7 Initial Training Network
``Invisibles''). I.T.\ acknowledges support by the Alexander von
Humboldt Foundation. Our results were achieved with high-performance
computing resources (Tier-0) provided by PRACE on CURIE TN
(GENCI@CEA, France) and SuperMUC (GCS@LRZ, Germany), and with
computing time on the IBM iDataPlex system \emph{hydra} of the RZG.

%%%%%%%%%%%%%%%%%%%%%%%%%%%%%%%%%%%%%%%%%%%%%%%%%%%%%%%%%%%%%%%%%%%%%%
%%%  Bibliography  %%%%%%%%%%%%%%%%%%%%%%%%%%%%%%%%%%%%%%%%%%%%%%%%%%%
%%%%%%%%%%%%%%%%%%%%%%%%%%%%%%%%%%%%%%%%%%%%%%%%%%%%%%%%%%%%%%%%%%%%%%

%%%%%%%%%%%%%%%%%%%%%%%%%%%%%%%%%%%%%%%%%%%%%%%%%%%%%%%%%%%%%%%%%%%%%%
\end{document}